\documentclass{nature-2}

\usepackage{graphicx}
\usepackage{verbatim}
\usepackage{subfigure}
\usepackage{tikz}
\usepackage{amsmath}
\usepackage{amssymb}
\usepackage{array}
\usepackage{tabularx}
\usepackage{multirow}
\usepackage{multicol}
\usepackage{rotating}
\usepackage{color}
\usepackage{layout}
\usepackage{epstopdf}
\usepackage{afterpage}
\usepackage[english]{babel}
\usepackage{authblk}
\usepackage{ulem}

\renewcommand{\thefigure}{{\bf \arabic{figure}}}

\title{Dynamics of beneficial epidemics}

\author{Andrew Berdahl$^{1}$, Christa Brelsford$^{1,2}$, Caterina De Bacco$^{1}$, Marion Dumas$^{1}$, Vanessa Ferdinand$^{1}$, Joshua A. Grochow$^{1}$, Laurent H\'ebert-Dufresne$^{1}$, Yoav Kallus$^{1}$, Christopher P. Kempes$^{1}$, Artemy Kolchinsky$^{1,3}$, Daniel B. Larremore$^{1}$, Eric Libby$^{1}$, Eleanor A. Power$^{1}$, Caitlin A. Stern$^{1}$ \& Brendan D. Tracey$^{1,3}$}

\begin{document}

\maketitle

\begin{affiliations}
 \item Santa Fe Institute, Santa Fe, NM 87501, USA
 \item Arizona State University, Tempe, AZ 85281, USA
 \item Massachusetts Institute of Technology, Cambridge, MA 02139, USA
\end{affiliations}

\begin{abstract}
Pathogens can spread epidemically through populations. Beneficial contagions, such as viruses that enhance host survival or technological innovations that improve quality of life, also have the potential to spread epidemically. How do the dynamics of beneficial biological and social epidemics differ from those of detrimental epidemics? We investigate this question using three theoretical approaches. First, in the context of population genetics, we show that a horizontally-transmissible element that increases fitness, such as viral DNA, spreads superexponentially through a population, more quickly than a beneficial mutation. Second, in the context of behavioral epidemiology, we show that infections that cause increased connectivity lead to superexponential fixation in the population. Third, in the context of dynamic social networks, we find that preferences for increased global infection accelerate spread and produce superexponential fixation, but preferences for local assortativity halt epidemics by disconnecting the infected from the susceptible. We conclude that the dynamics of beneficial biological and social epidemics are characterized by the rapid spread of beneficial elements, which is facilitated in biological systems by horizontal transmission and in social systems by active spreading behavior of infected individuals. 
\end{abstract}

Epidemiology has traditionally focused on the spread of harmful contagions, including human viruses such as influenza or dengue fever\cite{salathe2010high, yang2015forecasting, barmak2016modelling}, chytrid fungus in frogs\cite{pounds2006widespread}, and bacterial wilt in beans\cite{osdaghi2015first}. The serious consequences of detrimental epidemics drive their study, but \textit{beneficial} elements could also spread contagiously, and comparatively little is known about their dynamics\cite{shen2009challenge}. The social sciences have
studied the spread of beneficial behaviors, such as good health practices\cite{centola2010spread} or adoption of agricultural technology\cite{maertens2013measuring}, but such efforts have stopped short of a
unified mathematical framework that captures the wide range of observed phenomena and connects them with these epidemiological processes.

Beneficial epidemics involving the spread of viruses, plasmids, genes, and microbes, have been identified in biology, but their dynamics are poorly
characterized.
Beneficial viruses, which occur in both unicellular and multicellular organisms\cite{fraile2016environment}, can enhance survival\cite{roossinck2011good} or even be essential to the survival of the host\cite{stoltz2009virology}. 
Many beneficial genetic elements have been identified that spread horizontally among unicellular organisms\cite{xu2014densovirus, pradeu2016mutualistic, weeks2004increased, haine2008symbiont, brumin2011rickettsia, hedges2008wolbachia, brownlie2009symbiont}, but their large-scale dynamics are not well understood. Thus, while the dynamics of biological beneficial epidemics have been investigated in a small number of specific, unstructured populations\cite{van2001dangerous, xu2014densovirus, jiu2007vector}, many open questions remain\cite{shen2009challenge, haine2008symbiont, xu2014densovirus}. 

\begin{table*}[t]
    \begin{center}
    \small
     \begin{tabular}{| p{2.8cm}| p{4.3cm} | p{4.3cm} | p{4.3cm} | }
     \hline
     & {\bf  (Model 1)} & {\bf (Model 2)} & {\bf (Model 3)} \\
     \hline
         & \includegraphics[width=.25\textwidth]{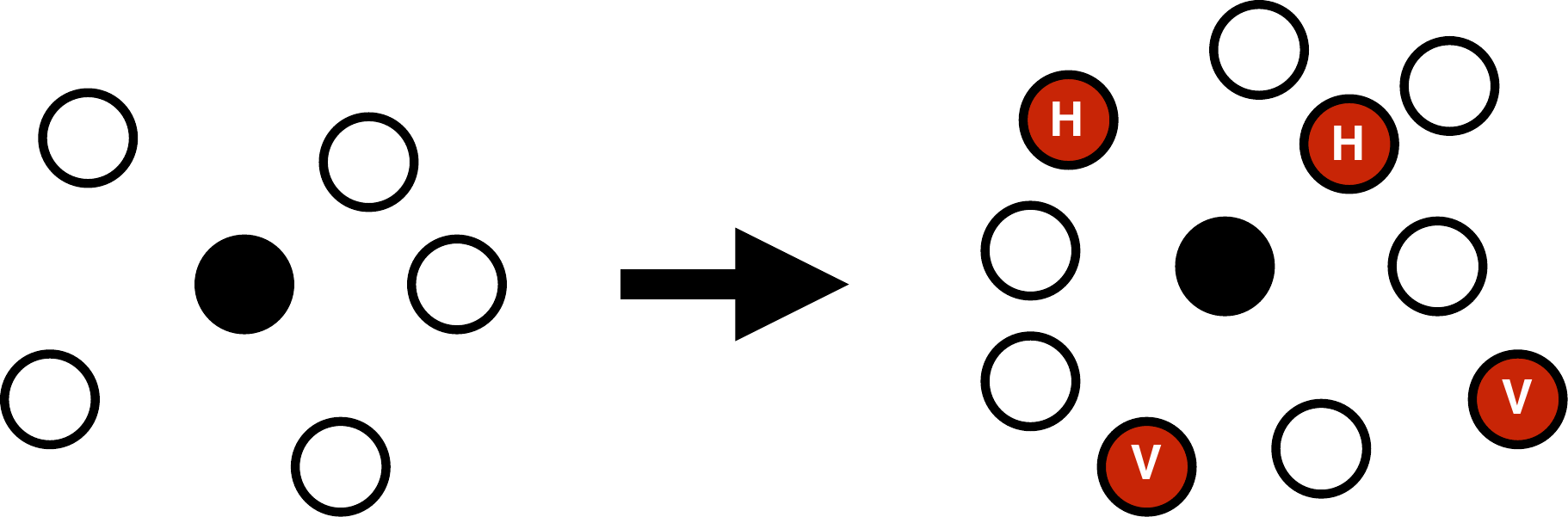} & \includegraphics[width=.25\textwidth]{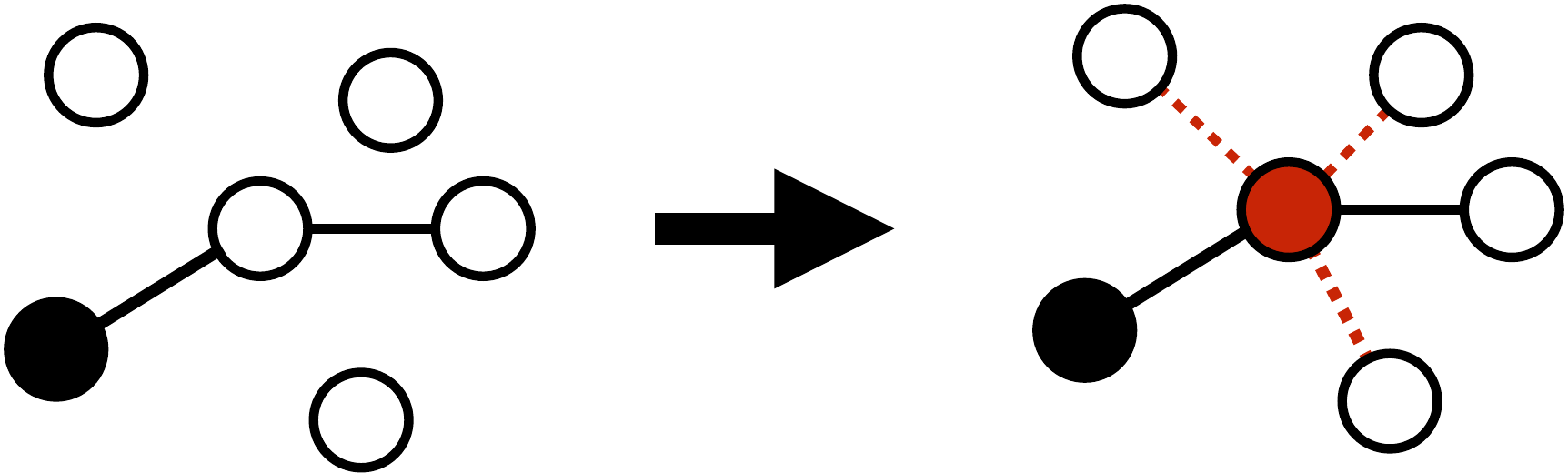}  & \includegraphics[width=.25\textwidth]{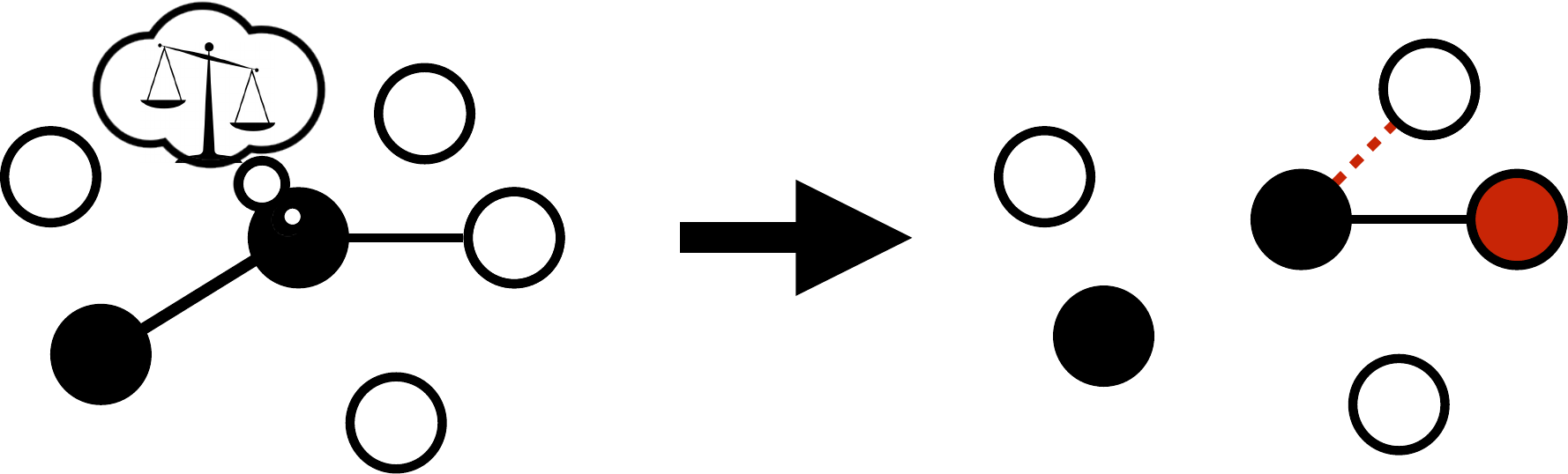}  \\
     \hline 
     {\bf Mode of Benefit} & fitness & implicit via connectivity & explicit utility\\
     \hline
   {\bf Population} & well-mixed & social & social \\ 
   \hline
   {\bf Transmission} & vertical and horizontal & horizontal & horizontal \\ 
   \hline
   {\bf Network rewiring} & none & fixed targeting & dynamic strategy \\ 
        
   \hline
   \multicolumn{1}{c}{} & \multicolumn{1}{c}{} & \multicolumn{1}{c}{} & \multicolumn{1}{c}{}  \\
    \end{tabular}
    \caption{Comparison of bene models. In the three schematics, the black arrow represents one increment of time, the black circles are infected individuals, the red circles are newly infected individuals, and the open circles are susceptible individuals. In Model 1, \textit{H} represents infection from horizontal transmission and \textit{V} from vertical transmission. In Model 2, dashed red lines indicated new social connections and solid black lines indicated existing connections. The same holds for Model 3, with the addition of strategic rewiring, which includes both adding new links and severing certain existing ones.}
\label{model-overview}
    \end{center}
   \end{table*}

In the behavioral sciences, social epidemics of behaviors, ideas, and technologies have long been studied (e.g. Ref.~\citenum{rogers2010diffusion}, first published in 1962). Examples include the adoption among humans of new agricultural technologies\cite{maertens2013measuring}, linguistic variants\cite{enfield2008transmission}, and social movements\cite{siegel_2008,snow_social_1980}, and extend to the acquisition of new feeding techniques among blue tits\cite{lefebvre1995opening} and humpback whales\cite{allen2013network}. In particular, studies have considered the influence of static network topology on spreading dynamics\cite{shipan_volden_2008, siegel_2008, simmons_elkins_2004, snow_social_1980,
conley_social_2001, bandiera2006social, maertens2013measuring},
but the simultaneous dynamics of networks and spreading remain largely unexplored.

Studying the dynamics of beneficial epidemics requires a clear definition of benefit. Here, by beneficial, we always mean beneficial to individual hosts, yet each model requires a specific definition of benefit relevant to that model. For example, in evolutionary contexts, benefits may be defined as conferring reproductive fitness, while in social contexts, benefits may be defined as increasing social influence or personal utility. By using a natural definition of benefit in each model, we are able to study the factors that unify dynamics of beneficial epidemics in a range of contexts. In contrast, one example of a genetic element that is horizontally transmitted but is not a beneficial epidemic is a lytic phage that induces infected cells to rupture. Because these phages harm the host there is the potential for a co-evolutionary arms race -- a dynamic not seen in beneficial epidemics.

We investigate epidemics of beneficial elements, which we call {\it benes}, in three contexts (Table \ref{model-overview}). In the context of population genetics, the relevant concept of benefit is simply reproduction. Therefore, we examine a bene that improves fitness and is transmitted both vertically and horizontally. In the context of behavioral epidemiology, a relevant benefit must be manifested within the same generation and affect social behavior. We thus analyze a bene that causes the formation of new network links that preferentially target uninfected nodes. In the context of dynamic social networks with individual agent preferences, the concept of benefit must incorporate the opinions of individuals about what is beneficial to them. In this context, we investigate a family of benes that incite individuals to form new social ties and break existing ones, causing a wide variety of transient and steady-state behaviors. Across these contexts, we find that the spreading dynamics of benes is qualitatively different than in traditional epidemics.

\section*{Model 1: Epidemics with fitness benefits}
In biological systems, beneficial mutations increase the reproductive fitness of an organism, increasing the number of offspring the host leaves in subsequent generations. Here we consider a beneficial sequence of genetic material that also spreads horizontally through the population, and contrast its spread with that of a beneficial element that is only transmitted vertically. 

We consider two types of individuals: those infected by the bene, $I$, and those uninfected by the bene and therefore susceptible, $S$.
The bene is assumed to increase the reproductive rate of infected individuals. On its own, a growth rate advantage would cause the infected population to eventually outnumber the susceptible population, but in this model, the bene can also spread horizontally between individuals, and so $S$ entities are also converted into $I$ entities within the same generation (see Supplemental Text for consideration of a fixed population model). We assume that the bene is transmitted across generations with probability $p$. The growth and infection processes are captured by the time evolution of the $S$ and $I$ population sizes:
\begin{equation}
	\dot{S} = S - \beta SI\text + (1 + s)(1-p) I, \qquad
	\dot{I} = (1 + s)p I + \beta SI\ \text,
	\label{EvoExpand}
\end{equation}
where $\beta$ is the infection rate. This basic model is similar to S-I models and work done previously (e.g. Ref.~\citenum{anderson1992infectious}).

In the absence of horizontal transmission ($\beta=0$) and assuming vertical transmission is perfect ($p=1$), \eqref{EvoExpand} can be solved analytically: $S(t)=S(0) e^{t}$ and $I(t)=I(0) e^{(1+s)t}$. The proportion of the population infected at time $t$ is therefore $1/(1+Z_0 e^{-s t})$, where $Z_0=S(0)/I(0)$, and as expected, the fraction of uninfected individuals shrinks exponentially. However, with horizontal transmission, $\beta > 0$, the population is taken over by infected individuals much faster than without horizontal transmission. Consider a slight variant of \eqref{EvoExpand}, where susceptible individuals that become infected through horizontal transfer are removed from the susceptible population, but do not add to the infected population. Under this assumption, the number of infected individuals is the same as without horizontal transfer and the number of susceptible individuals becomes
\begin{equation}
	S(t)=S(0) e^{t-\frac{\beta}{1+s} e^{(1+s)t}+\frac{\beta}{1+s}}\text.
\end{equation}
The proportion of the population still uninfected at time $t$ is therefore 
\begin{equation}	
    Z(t)=\left [ 1+Z_0^{-1} e^{\frac{\beta}{1+s}(e^{(1+s)t}-1)} e^{s t} \right ] ^{-1}\text,
\end{equation}
exhibiting a superexponential decay, and this is an upper bound on $Z(t)$ for the full model (Eq.~(\ref{EvoExpand})).

If vertical transmission is imperfect ($p<1$) then the infected $I$ population continually generates susceptible individuals. The $S$ population approaches a steady state of $S^*=(1-p)(1+s)/\beta$ as $t \to \infty$. If we assume that the infected population is initially small, then the system has three dynamical regimes (see Supplemental Figure S1 for example). First, when $\beta SI\ll S,I$, the population is so dilute that horizontal transmission events are rare, and both populations grow exponentially such that $S \propto e^t$ and $I \propto e^{(1+s)p t}$. As $\beta SI$ increases there is a sharp transition phase where susceptible individuals are rapidly infected. This leads to the final phase where the $S$ population approaches the steady state and the $I$ population grows exponentially at a rate $I \propto e^{(1+s) t}$ that is independent of the vertical transmission probability. Of the three dynamical regimes, the transition phase is the only one with  potential for super exponential dynamics since the $I$ population is moving from one exponential growth rate to another, larger one.

\section*{Model 2: Epidemics with connectivity benefits}\label{epidemiology}
In the previous section, we considered benes whose beneficial effect occurs across generations in a well-mixed population, but in a heterogeneous population, the spread of a bene can be affected in the same generation by a change in the social behavior
of infected individuals. Many modes of benefit manifest indirectly in an increased social connectivity of infected individuals: increased energy allowing more social connections, increased social desirability attracting new contacts, or conscious desire to spread the bene. Initially, we do not explicitly model the underlying benefit, but only consider its indirect effect on the network
of social contacts in a population. We obtain a preliminary view of the spreading dynamics we expect for such benes.
A more explicit consideration of how a contagion's benefit might induce a change in social behavior is the basis of the model analyzed in the next section.

Harmful contagions can also induce behavioral effects that increase its spread, and the interplay between an infectious disease and changes in the underlying network structure has been studied at great length\cite{gross2006epidemic, risau2009contact, funk2010modelling, marceau2010adaptive, volz2011effects, althouse2014epidemic, leventhal2015evolution,scarpino2016prude}. However, because the contagion is detrimental to the host, there is usually a tension between behavior of infected individuals, affected by the contagion to try to increase its spread, and that of the uninfected population, attempting to limit it. The spread of a purely beneficial contagion would not involve this tension and we expect different spreading dynamics.

We consider an ``SIS'' model where nodes can be either infected or susceptible, and may transition from either state to the other.
Susceptible nodes are infected at a transmission rate $\beta$ by each of their infected neighbors and infected nodes recover at rate $r$ to become susceptible.
We suppose that the consequence of the bene is to generate $\Delta$ new links upon infection and remove the same amount upon recovery.
We also suppose the targets of new links to be chosen either randomly from all nodes in the network or preferentially chosen
from susceptible nodes (disassortative). 
This preference is modeled with the parameter $\alpha$ that denotes the assortative bias. When $\alpha = 0$, susceptible nodes are always selected as the target for new links by infected nodes. When $\alpha = 1$, there is no bias, and targets are chosen uniformly from the population. In between, susceptible nodes are preferentially targeted, but links of both types can be created.

Let $S$ and $I$ denote the fraction of nodes currently susceptible and infected, respectively, such that $S+I=1$.
Let $[SI]$ be the number of edges between $S$ and $I$ nodes normalized by the total population size,
and so on for $[SS]$ and $[II]$. Following the methods of Refs.\cite{gross2006epidemic, risau2009contact} for networks with Poissonian degree distribution of mean $k_0$, we can write the differential equations governing this process as follows.
\begin{align}\label{eq:ode-inst}
\dot{I}&=-\dot{S}= \beta [SI] -rI \notag\\
\dot{[SS]}&= -2\beta \,([SS][SI]/S)+r [SI] \frac{k_I-\Delta}{k_I} \notag\\
\dot{[SI]}&= 2\beta \,([SS][SI]/S)  - \beta([SI]^2/S) -\beta[SI] -r[SI] +\\
&\quad\beta[SI]\Delta \frac{S}{S+I\alpha} +2\,r\,[II] \frac{k_I-\Delta}{k_I}  \notag\\
\dot{[II]}&= \beta([SI]^2/S) + \beta[SI]+\beta\,[SI]\, \Delta \frac{I\alpha}{S+I\alpha} -2r [II] \text, \notag
\end{align}

The average degree of an infected node is $k_{I}=(2[II]+[SI])/I$. The critical transmission rate for the bene to spread epidemically is then $\beta_c=r/\left(\tau+\sqrt{k_0+\tau^2}\right)$, where $\tau=\tfrac12(k_0+\Delta-1)$. When $\beta<\beta_c$, any small infection dies out and the only stable state is when the entire population is susceptible ($S=1$). When $\beta>\beta_c$ the $S=1$ equilibrium becomes unstable and an arbitrarily small infected population will grow to an extensive size. 

In a static Poissonian network, the epidemic threshold $\beta_c$ is simply $r/k_0$. Notice that while we recover this result in the limit $\Delta \rightarrow 0$, our critical transmission rate is not simply that of a Poissonian network with average degree $k_0 + \Delta$. On the one hand, the degree distribution of infected node is of smaller variance than a Poisson distribution, which \textit{raises} the epidemic threshold. On the other hand, there is a feedback between the expected epidemic size and the average degree of the network which \textit{lowers} the epidemic threshold. Our steady state analysis is illustrated in Figure S3 as a function of model parameters.

We saw in the biological model that the steady state proportion of uninfected individuals can decrease toward zero superexponentially due to a combination of a fitness disadvantage and horizontal transmission. Using the present model, we find that such a superexponential decrease can also occur due to a combination of horizontal transmission and targeted link generation.

If new links are perfectly targeted at susceptible individuals ($\alpha=0$), then as long as more than one link on average is generated per infection ($\Delta \ge 1$), the susceptible population shrinks double-exponentially, that is, $dS(t)/dt = -x(t) S(t)$, where $dx(t)/dt = \beta (\Delta-1) x(t)$.
On the other hand, if $\Delta<1$, then even if new links are perfectly targeted, the rate at which $S$ decreases itself decreases exponentially. 

Suppose that infected individuals imperfectly target susceptible individuals. Then, as $I$ becomes much larger than $S$, even a small $\alpha >0$ causes most new links to be made toward already infected individuals.
Effectively, this is equivalent to $\Delta\rightarrow 0$. The rate at which $S$ decreases will decrease exponentially, as in a standard epidemic process.
We expect two phases in the final spreading dynamics: at first, when $\alpha\ll S$, the behavior will be as if $\alpha=0$, with double-exponential decay of the susceptible population size (assuming $\Delta>1$). However, eventually $S$ becomes smaller than $\alpha$, and the system acts as if no extra $SI$ links are generated.

\begin{figure}
	\centering
	\includegraphics[width=1.0\linewidth]{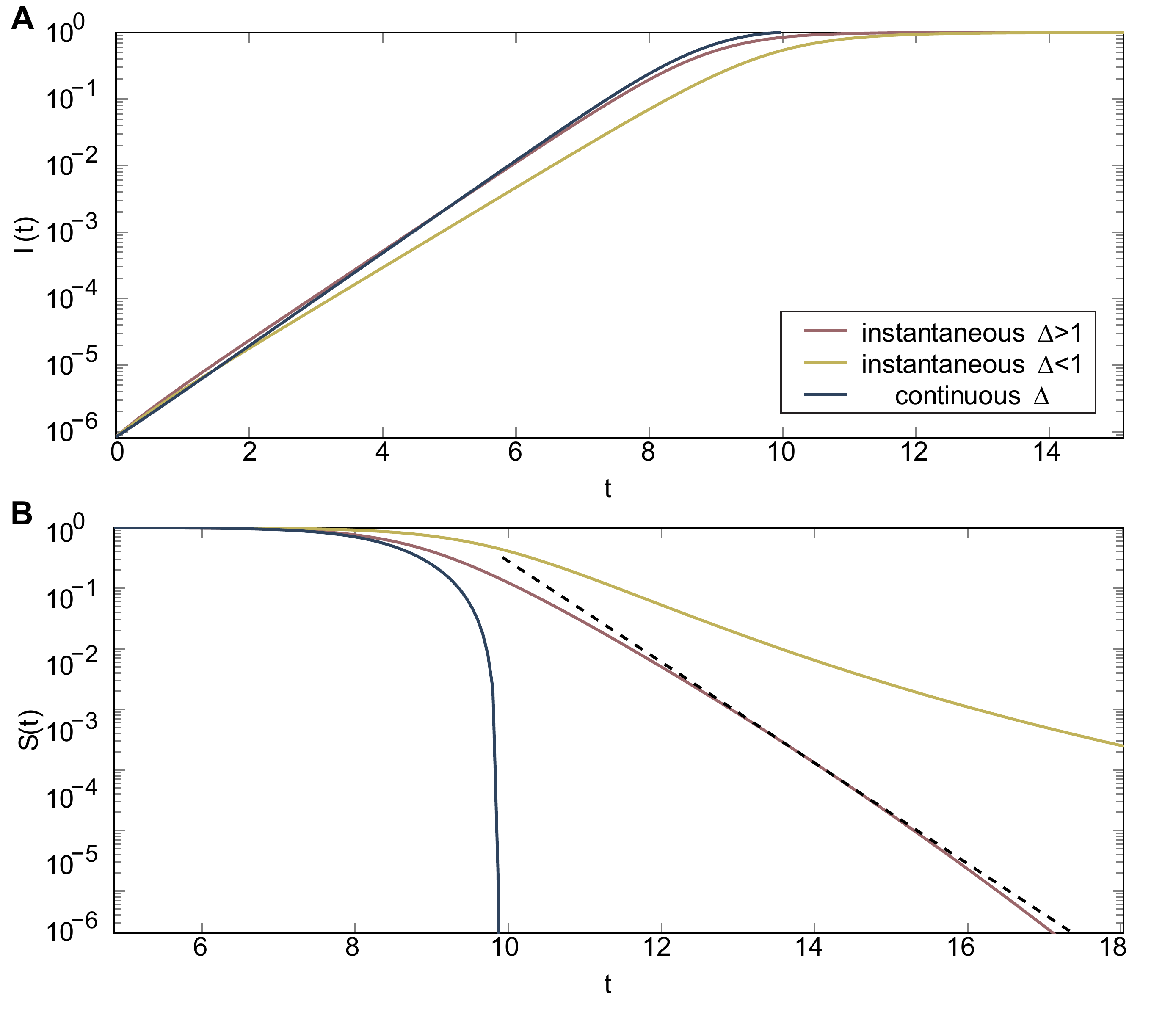}
	\caption{\linespread{1.0} \selectfont{}  Perfect targeting $(\alpha=0)$ initial and final dynamic regime. (A) The fixation dynamics are shown for both the instantaneous and continuous link generation models. In the continuous model the convergence to complete fixation is in finite time. For the instantaneous model, complete fixation is approached faster than exponentially (when $\Delta>1$, dotted straight line for comparison) or is not approached even for very long times (when $\Delta<1$). (B) The initial dynamics is given for the same cases. In all these plots the recovery rate is $r=0$ and the transmission rate $\beta=1/2$, the initial degree $k_0=3$.}
	\label{fig:epilogSI}
\end{figure}

In the Supplementary Information, we analyze a similar model where the extra connectivity accrues throughout the time an individual is infected. We then find that the infection always reaches fixation in finite time. Analytic results are summarized in Table~\ref{tab:summary} and illustrated in Figure~\ref{fig:epilogSI}.

   \begin{table}[tb]
 \caption{Dynamics of benes with connectivity benefits}
 \label{tab:summary}
 \begin{center}
 \begin{tabular}{c|cccc} 
 \hline
 & \multicolumn{2}{c}{instantaneous $\Delta$} & \multicolumn{2}{c}{continuous $\Delta$} \\ 
	 & $\alpha = 0$ & $\alpha > 0$ & $\alpha = 0$ & $\alpha > 0$ \\
 \hline
 epidemic threshold & \multicolumn{2}{c}{$\tfrac{\beta_c}{r} = \tfrac{1}{\tau +\sqrt{k_0+\tau^2}}$} & \multicolumn{2}{c}{---}\\
  & \multicolumn{2}{c}{$\tau=\tfrac12(k_0+\Delta-1)$} & &\\
 early time & \multicolumn{2}{c}{exp. growth; const. rate} & \multicolumn{2}{c}{exp. growth; variable rate} \\
 fixation & $e^{-e^{t}}$  & $e^{-e^{-t}}$ & $(t^*-t)^2$  & $e^{-t}$ \\
 \hline
 \end{tabular}
 \end{center}
\end{table}

The behaviors observed in our epidemiological models differ dramatically from classic spreading dynamics. In short, while the addition of the connectivity benefit has a straightforward impact on the epidemic threshold and the early time spread, the fixation dynamics are sensitive to both how these new links are created and to whom. Fixation dynamics refer to the spread of the epidemic in the endemic regime, where we would expect detection of benes to be most likely. Our predictions for possible behaviors in this regime provide us with potential signals for the detection of benes in empirical data. Moreover, the sensitivity of these predictions to individual preference (i.e., $\alpha$) suggests the need for frameworks to incorporate the costs and benefits of social targeting.

\section*{Model 3: Epidemics with utility benefits}\label{agency}

The previous section considered how a bene spreads when its implicit benefit manifests in increased connections. Here, we consider the case when a bene has explicit consequences for an individual's utility. We call an infectious trait a bene if becoming infected leads to an increase in utility. We consider how the utility conferred by the infection leads individuals to rewire strategically so as to influence infection dynamics and thereby increase their future expected utility. For example, if infected individuals can increase utility by growing the size of the infected population, they can `proselytize' and spread the trait by seeking out new social connections to the susceptibles; on the other hand, if the infected gain utility from only being connected to other infected, the opposite rewiring dynamic can take hold.

We consider an epidemiological model in which both infected and susceptible individuals rewire their connections based on a utility function. The utility function reflects preferences for local conformity versus global spreading of the infection.  Infected and susceptible individuals' utility functions are indicated by $U_I(I_n,S_n,I_g)$ and $U_S(I_n,S_n,I_g)$ respectively, where $I_n$ is the number of infected neighbors, $S_n$ is the number of susceptible neighbors, and $I_g$ is the total number of infected individuals in the global population.

Importantly, individuals rewire based on their predictions of how their future expected utility will change due to epidemic spreading dynamics. In making predictions, individuals only make use of knowledge of their direct connections (and not, for example, connections between their neighbors).  We use $P_I$ and $P_S$ for the predicted expected utility of infected and susceptible individuals. As before, transmission is assumed to be a simple contact process with rate $\beta$. 

We show that different preferences for local and global infections lead to different dynamical regimes.
Here we assume that the utility functions are linear: $U_I(I_{n},S_{n}, I_g) = a_I I_n + b_I S_n + c_I I_g$ and 
$U_S(I_{n},S_{n}, I_g) = a_S I_n + b_S S_n + c_S I_g$, where $a_I$ and $a_S$ are parameters specifying the utility of one additional infected neighbor, $b_I$ and $b_S$ specify the utility of one additional susceptible neighbor, and $c_I$ and $c_S$ specify the utility of increasing the number of infected individuals in the global population by one. We also define $d_I = a_I-b_I$ and $d_S = a_S-b_S$, the utilities of swapping a susceptible neighbor for an infected one.

Infected individuals' predictions account for the probability that they will infect some number $X = 1,\dots,S_n$ of their susceptible neighbors. Assuming a well-mixed population, $X$ is distributed as a binomial, $X \sim B(S_n, \beta)$, giving
\begin{align*}
P_I(I_n,S_n,I_g) & = \mathbb{E}_X[ U_I(I_n + X, S_n -X, I_g + X)] \\
    & =  U_I(I_n , S_n, I_g) + (d_I+c_I)\beta  S_n \text.
\end{align*}
Susceptible individuals account for the probability that they become infected by at least one of their infected neighbors
\begin{align*}
P_S(I_n,S_n,I_g) &= (1-(1-\beta)^{I_{n}}) U_I(I_{n},S_{n},I_g+1) + (1-\beta)^{I_{n}} U_S(I_{n},S_{n},I_g) \\
&\approx \beta I_{n}\left(U_{I}\left(I_{n},S_{n},I_{g}\right)+c_I\right)+\left(1-\beta I_{n}\right)U_{S}\left(I_{n},S_{n},I_{g}\right)\text,
\end{align*}
where we use $(1-\beta)^{I_n} \approx 1-\beta I_n$ assuming $\beta I_n \ll 1$.

The parameters $a_I, \dots, c_S$ determine individuals' rewiring behavior.  For an infected individual, the change in predicted utility from disassortative rewiring is
\begin{equation}
\nonumber
P_I(I_n -1, S_n+1, I_g) - P_I(I_n, S_n, I_g) = -d_I+\left(d_I+c_I\right)\beta
\end{equation}
Assortative rewiring is chosen when: 
\begin{equation}\label{eq:assortative-infected}
d_I \geq \left(d_I+c_I\right)\beta
\end{equation}
while disassortative rewiring is chosen otherwise. For infected individuals, the predicted utility of each rewiring strategy does not depend on the state of the population: either the assortative or the disassortative regime will hold for all infected individuals at all times.

For a susceptible individual, the change in predicted utility from disassortative rewiring is
\begin{equation}
\nonumber
d_S+\beta\left[(a_I-a_S)(I_n-S_n) + 2(b_I-b_S)S_n + (c_I-c_S)I_g+c_I\right]
\end{equation}
with the negative for assortative rewiring. Thus, for susceptible individuals, the predicted utility of each rewiring strategy depends on the state of the population, and not just the parameters. The assortative rewiring will be chosen when
\begin{multline}
(a_I-a_S + d_I-d_S)I_n +  (b_I-b_S)S_n + (c_I-c_S)I_g  
< -\frac{d_S}{\beta}-c_I + (d_I-d_S)\label{eq:assortative-susceptible}
\end{multline}
and disassortative otherwise.

This framework tells us how individual preferences translate into assortative or disassortative rewiring behavior for infected and susceptible individuals in the course of an epidemic. This allows us to formulate dynamics similar to those presented in the earlier sections (see Supplemental Text C for the general derivation), but now derived from the preferences and predictions of individuals. By combining epidemiological modeling with strategic rewiring, our framework could be used to analyze social movements, the spread of technologies, and strategic rewiring dynamics in detrimental infections (e.g., in which individuals rewire to avoid infection\cite{fenichel2011adaptive,delvalle2005}).

We now consider three illustrative cases, corresponding to three different sets of parameter values. In the \textit{evangelizers} case, the utility of infected individuals increases when the infection spreads globally, corresponding to $c_I=1$ (non-specified parameters are $0$). In this case, infection causes an increase in utility and is thus a bene. Based on \eqref{eq:assortative-infected} and \eqref{eq:assortative-susceptible}, infected as well as susceptible individuals rewire disassortatively, tempted by the possibility of increasing global spread. In the \textit{cool kids} case, all individuals prefer increasing the number of infected neighbors (the ``cool'' kids) and decreasing the number of susceptible neighbors (the ``uncool'' kids), corresponding to $a_I=a_S=1$ and $b_I=b_S=-1$. In this case, infection confers no direct net utility change. Yet, infected individuals rewire assortatively, while susceptibles rewire disassortatively. Finally, in the \textit{snobs} case, infected individuals prefer to be connected to other infected individuals, while susceptibles are indifferent, corresponding to $a_I = 1$, $b_I=-1$. In this case, whether or not infection causes an increase in utility depends on an individual's neighborhood. As a result, susceptibles exhibit a complicated behavior: they switch from assortative to disassortative behavior at a particular cutoff number of infected neighbors $I_n> \frac{S_n + 2}{3}$. As long as susceptible neighbors are numerous, susceptible individuals are assortative to avoid the risk of an infection, which would put them at odds with their neighborhood.  Once infected neighbors become sufficiently numerous, the susceptible become disassortative to have the chance to become infected and conform.

These three cases are motivated by previously studied social processes. The evangelizers case can be seen as a model of explicit recruitment in a social movement~\cite{snow_social_1980, olson2009logic}. The cool kids case reflects the transmission of an idea through a group, where infected individuals' assortativity results in the formation of cliques (e.g., the anticonformity copying modeled in\cite{boyd1988culture}). The snobs case is related to models of segregation\cite{schelling1971, clark_and_fossett_2008} where potentially asymmetric and conflicting preferences for assortativity exist.

We simulate the dynamics of this model with ODEs (Supplemental Text C) similar to the ones in the previous sections. In addition to contact spreading dynamics, assortative and disassortative rewiring is performed according to the rules described above. We assume a well-mixed compartmental model with $N$ individuals and $E$ edges, in which we track the proportion of infected individuals $I$, and the proportion of $[II]$, $[SI]$, and $[SS]$ links. Neighborhoods (i.e. values of $I_n$ and $S_n$) are assumed to be drawn with replacement from these compartments.

\begin{figure*}
\centering
\includegraphics[width = 1.0\linewidth]{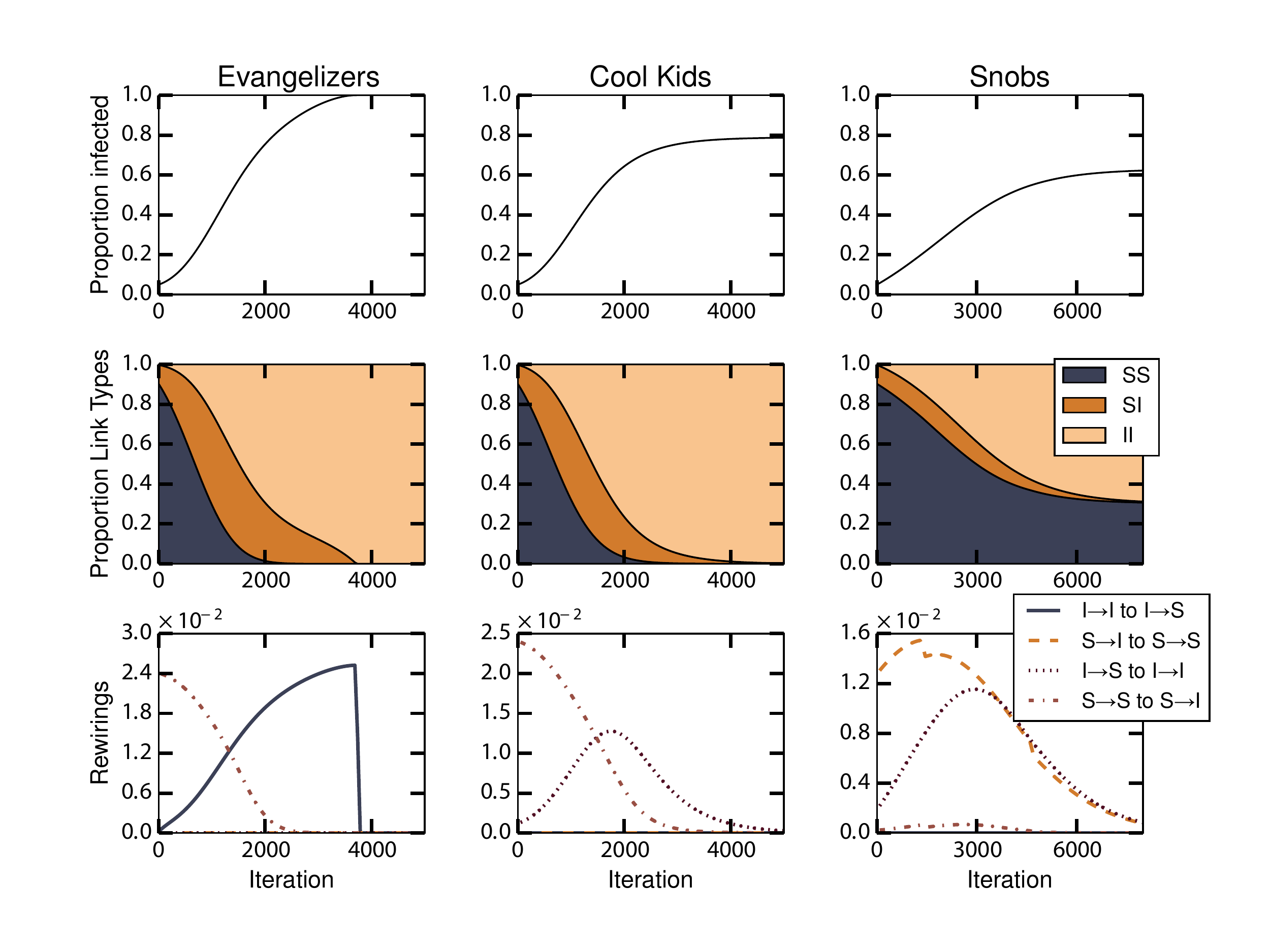}
\caption{\linespread{1.0} \selectfont{} Dynamical regimes in the social epidemiological model. The dynamical regimes arising for the three different cases. The first row shows the number of infected individuals over time. The second row shows proportion of $[II]$, $[SI]$ and $[SS]$ over time. The third row shows the rate of different rewirings, where for example, `I $\rightarrow$ I to I$\rightarrow$S' indicates the rewiring by an infected individual from an infected to a susceptible. Here $N=1000$, $E=4000$, $I_{init} = .05$, and $\beta = 5\times 10^{-4}$.}
\label{fig:weirdresults}
\end{figure*}

Figure \ref{fig:weirdresults} shows the dynamics for each case. The evangelizers case exhibits the same superexponential fixation dynamics as in the epidemic model with connectivity benefits (see Supplemental Text C).  The disassortative behavior of both infected and susceptible individuals speeds up the epidemic and drives the fixation dynamics.
In the cool kids case, the epidemic is incomplete. The susceptible rush to rewire to the infected, as shown by initially high rates of `S$\rightarrow$S to S$\rightarrow$I' rewiring, while the infected break ties with the susceptible, as seen by the increase in the rate of `I$\rightarrow$S to I$\rightarrow$I' rewiring once the number of infected individuals rises. These two behaviors compete, but once the infected are sufficiently numerous the latter dominates and the doors to the infected community close. In the resulting network, all connections are between infected individuals ($[II]=1$), susceptibles are isolated, and the epidemic halts.  
In the snobs case, the epidemic reaches even fewer individuals than in the previous case. While the infected rewire assortatively, the susceptible have a mix of strategies. The result is that the network divides into two completely disconnected components ($[SI]=0$ and $[II] + [SS] =1$), preventing the epidemic from reaching the whole population. We also implemented this model using an explicit agent-based simulation, which included effects of stochasticity, local network heterogeneity, and correlations of connectivity properties across the network. The results were qualitatively similar to the ODE model results reported here (Supplemental Text C and Figure S4).

\begin{figure}
\centering
\includegraphics[width = 1.0\linewidth]{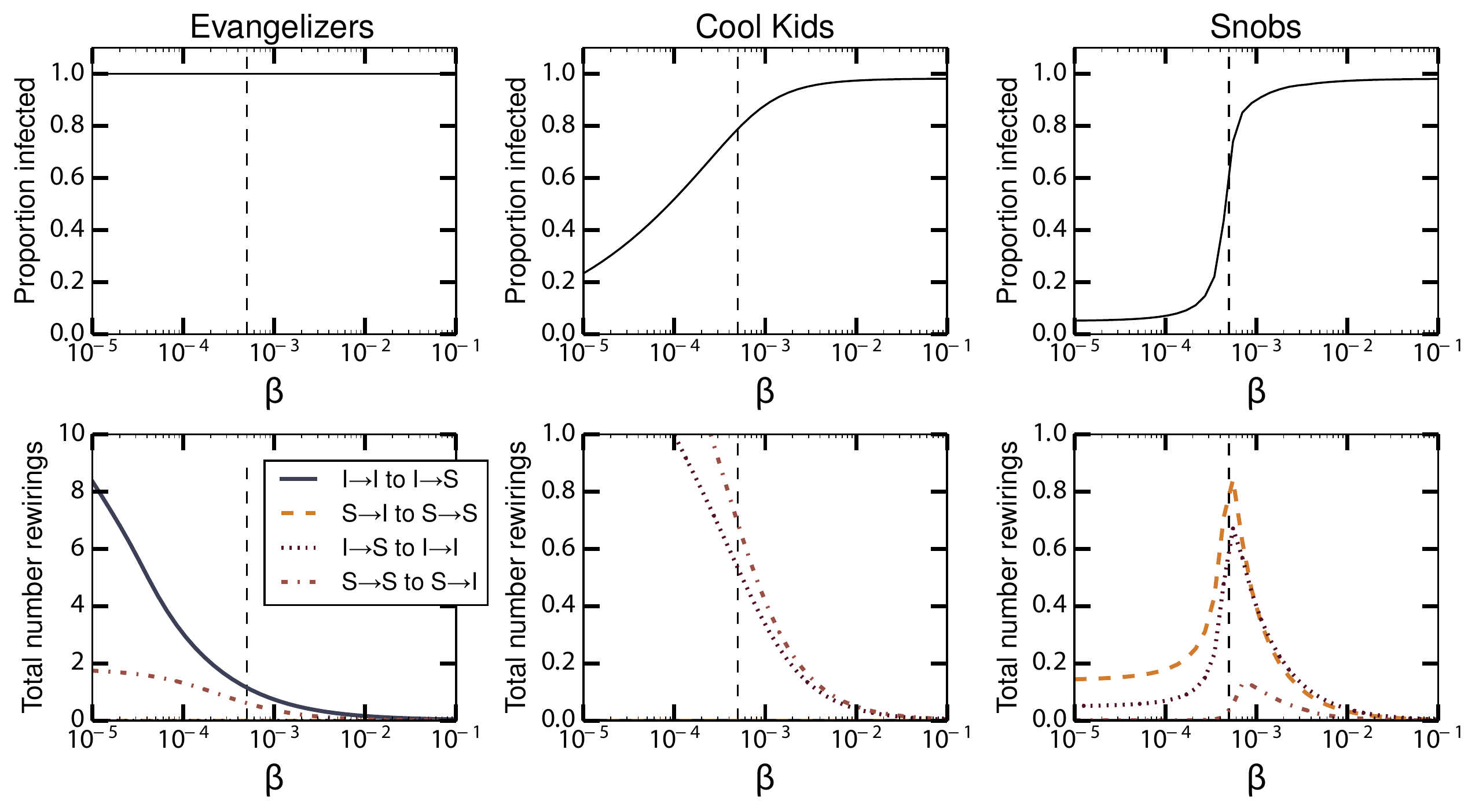}
\caption{\linespread{1.0} \selectfont{} Effect of transmission probabilities on epidemic spread and rewiring dynamics. Results of strategic rewiring epidemics with varying transmission rates $\beta$.  The top row shows the proportion of infected for the three cases discussed in the text.  The bottom row shows the cumulative rewiring performed (per edge). The value of $\beta$  used for Figure \ref{fig:weirdresults} is shown as a vertical dashed line.}
\label{fig:transmissionprob}
\end{figure}

The characteristics of these regimes vary with $\beta$. Figure \ref{fig:transmissionprob} shows the final reach of the epidemic and the cumulative rewiring for different values of $\beta$. For the evangelizers case, the epidemic always spreads to the entire population. For the cool kids case, the reach of the epidemic increases gradually with $\beta$, since faster spreading increases how many individuals get infected before the susceptible become disconnected.  For the snobs case, larger $\beta$ increases the reach of the epidemic, with a rapid transition from minimal spread to full spread once $\beta$ passes a critical threshold. In all these cases, the total amount of rewiring decreases when $\beta$ is sufficiently large, since rewiring stops once the epidemic has swept through the population.  Interestingly, the amount of rewirings for the snobs case also decreases with lower $\beta$, peaking around the critical threshold.  This occurs because when transmission is slow, equilibrium is reached very quickly: assortative rewiring by the infected and the susceptible quickly disconnects these two groups before the infection can spread. 

This model shows that strategic rewiring affects epidemic dynamics in multiple ways. When infected individuals benefit from increasing the global number of infections, it leads to an accelerating uptake and a much faster global spread than an epidemic without strategic rewiring. If instead the infected individuals tend to assort, the epidemic can be stalled as infected individuals entirely disconnect from susceptible ones. 

\section*{Discussion}
In this paper, we study the epidemics of beneficial contagions, benes.
We find that they can spread much more rapidly than pathogens that are traditionally studied.
We investigate benes in several distinct systems, both biological and social.
While the dynamics by which benes spread depend on the particular benefit conferred, we find commonalities across these systems.
One striking outcome is that all scenarios exhibit superexponential fixation in particular regimes. 

A prime example of superexponential behavior is found in our evolutionary model. Here, the bene confers a fitness advantage to infected individuals, and, in contrast to a standard positive mutation, the bene can also be transmitted across individuals within a generation.
One example of such a bene would be antibiotic-resistance cassettes\cite{Partridge757}, where bacteria acquire genes from neighboring cells that increase survival when exposed to antibiotics.
Our model shows that horizontal transfer of such elements, even when vertical transmission is imperfect, dramatically reduces the time required to fully infect the population. 

The importance of horizontal transfer in evolutionary processes led us to consider a bene which increases interactions between individuals in a network. An example is new technologies with network effects, like the file-sharing service Dropbox, that incentivize users to actively recruit new members. We use an ``SIS'' epidemiological model to analyze the effects of these added network links. We find that added connections change the epidemic threshold, allowing benes to break out despite lower transmissibility. We also find a much lower fixation time within the population. In fact, if new edges are added only with susceptible individuals, the bene sweeps the entire population in finite time. This result demonstrates that individual behavior is important in determining whether a beneficial epidemic occurs.

In the model of epidemics with utility benefits, individuals' behavior is based on preferences for the distribution of the infection in the local neighborhood and global population. They strategically rewire based on predictions about how their actions will increase utility. One example is the phenomenon called NIMBY\cite{shemtov2003social}, where individuals have a preference for global adoption of a technology but do not want it in their immediate neighborhood, e.g. wind turbines. As we show in three illustrative cases, variation in the strength of these preferences leads to different dynamical regimes. In one regime, displayed by the \textit{evangelizer} case, infected individuals rewire to susceptible ones, facilitating the bene's spread. Social movements instilling the desire to convert anybody, and not just acquaintances, will spread quickly.  In contrast, when individuals prefer to conform with their neighbors, as in the \textit{cool kids} and \textit{snobs} cases, assortative rewiring results in a disconnected network and a stalled epidemic. Thus, the outcome of an epidemic may reveal the mechanisms underlying its generating dynamics. 

In this paper we considered the dynamics of beneficial epidemics for certain biological and social systems. We investigated contagions that confer specific types of benefits related to fitness or social utility, but many other types of beneficial epidemics are possible. For example, one could combine elements of our three models so that changes in social networks have cross-generational effects. Alternatively, one could consider a contagion that is beneficial to one type of host but harmful to others. These more complex models may exhibit other interesting behaviors that differ from the more traditionally and extensively studied harmful epidemics. By differentiating between the dynamics of various types of epidemics, it may be possible to identify distinct signatures of epidemics and determine the type of contagion as it is spreading in real time. This line of research could ultimately improve our ability to prevent the spread of harmful epidemics and harness the power of beneficial epidemic dynamics to facilitate social change.

\textbf{Acknowledgments}. The authors acknowledge the insightful comments of two anonymous referees whose feedback greatly improved the work. We are grateful for generous financial support from the Miller Omega Program, and from the Santa Fe Institute, whom we thank for encouragement in developing and conducting the 72 Hours of Science experiment. We acknowledge M. Lachmann, V. Marceau, and J. Miller for helpful discussions. We especially thank M. Alexander, D. Bacon, B. Bertram, R. Butler-Villa, J. Dunne, J. Elliott, J. German, M. Girvan, M. Hamilton, D. Krakauer, J. Lovato,  N. Metheny, J. Miller, S. Redner, D. Reed, K. Serna, C. Shedivy, and H. Skolnik.

\bibliography{refs}

\clearpage
\newpage

\appendix

\renewcommand{\thefigure}{S\arabic{figure}}
\renewcommand{\thetable}{S\arabic{table}}
\setcounter{figure}{0}
\setcounter{table}{0}

{\Large Supplementary Information}

\section{Evolutionary model}
\subsection{Imperfect transmission numerical solution}

\begin{figure}
	\centering
	\includegraphics[width=0.6\linewidth]{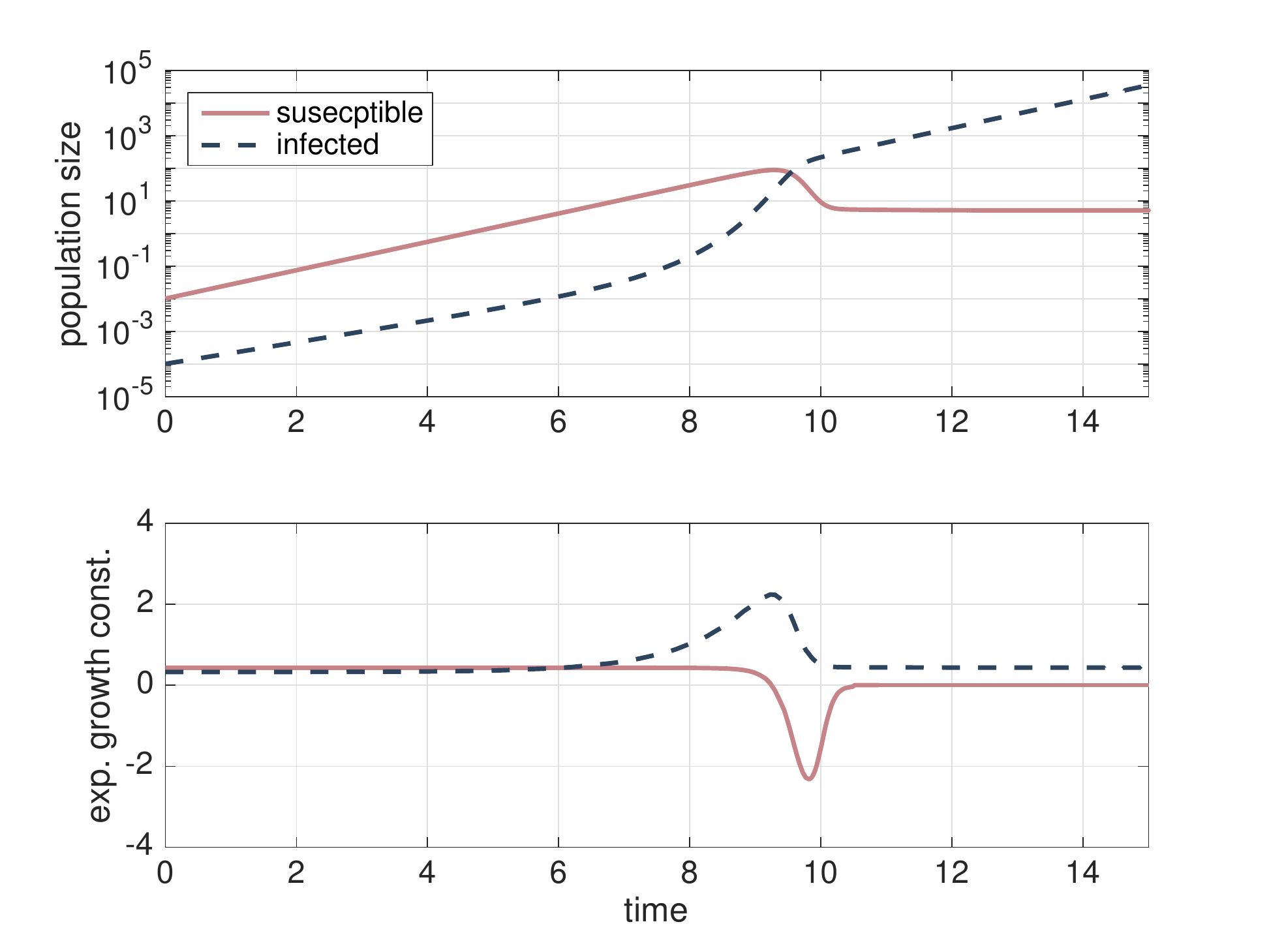}
	\caption{\textbf{Sample dynamics of model 1 with imperfect transmission.} Numerical solution of Eq. 1 with parameters $\beta=.05$, $s=.01$, and $p=.75$.}
    \label{fig:imperfecttrans}
\end{figure}

We solve the system of equations with imperfect transmission using parameters $\beta=.05$, $s=.01$, and $p=.75$. Thus, there is a small fitness benefit to the bene and it fails to transmits vertically with a probability of $.25$. Starting with an initial concentration of $.01$ for susceptible (red) and $.0001$ for infected (dashed), we see that there are distinct dynamical regimes (Figure \ref{fig:imperfecttrans}). First both population grow exponentially, then a rapid decline in the susceptible population, followed by steady growth for the infected and no growth for the susceptible population. The bottom panel shows the growth coefficient, i.e. the slope in log space for both populations. Only in middle regime is there the potential for super exponential dynamics.

\subsection{Fixed population size}
In the main paper, we considered a bene in an expanding population. Here, consider the same effect in a model with a finite population of size $N$. We use a previously published model of horizontal gene transfer\cite{novozhilov2005mathematical} which is easily generalized to other mechanisms of horizontal transmission of genetic elements found in multicellular organisms such as crustaceans and insects\cite{dupeyron2014horizontal}. This model (\eqref{EvoFinite}) distinguishes the effects of the transmission rate $\beta$ of the bene and the selective value $s$ of the bene. The number of $I$ entities is $n$ and the number of $S$ entities is $N-n$.  The model is based on a Moran process\cite{moran1958random} in which a birth/death process occurs in discrete time steps. The probability of having $n$ infected types at time $t$, $p_n(t)$, depends on the cumulative effects of the birth ($\lambda_n$) and death ($\mu_n$) rates (see \eqref{EvoFinite}). The birth rate ($\lambda_n$) includes actual births (the first term, which involves $s$) as well as horizontal gene transfer (the second term, which involves $\beta$). There are two stationary states of this model: either i) the entire population is un-infected (S), or ii) the entire population is infected (I). We solve for the stationary distribution of $p_n(t)$ starting with $p_n(0) = \delta_{n,1}$ (i.e all realizations start with a single infected individual ($n=1$)).
\begin{equation}
	\label{EvoFinite}
	\begin{aligned}
	\frac{dp_n}{dt} &= \mu_{n+1} p_{n+1} - (\lambda_n + \mu_n)p_n + \lambda_{n-1} p_{n-1} \\
	\lambda_n &= (1+s)n \frac{N-n}{N+1} + \beta \frac{n(N-n)}{N} \\
	\mu_n &= (N-n)\frac{n}{N+1}
	\end{aligned}
\end{equation}

\begin{figure}
	\centering
	\includegraphics[width=0.6\linewidth]{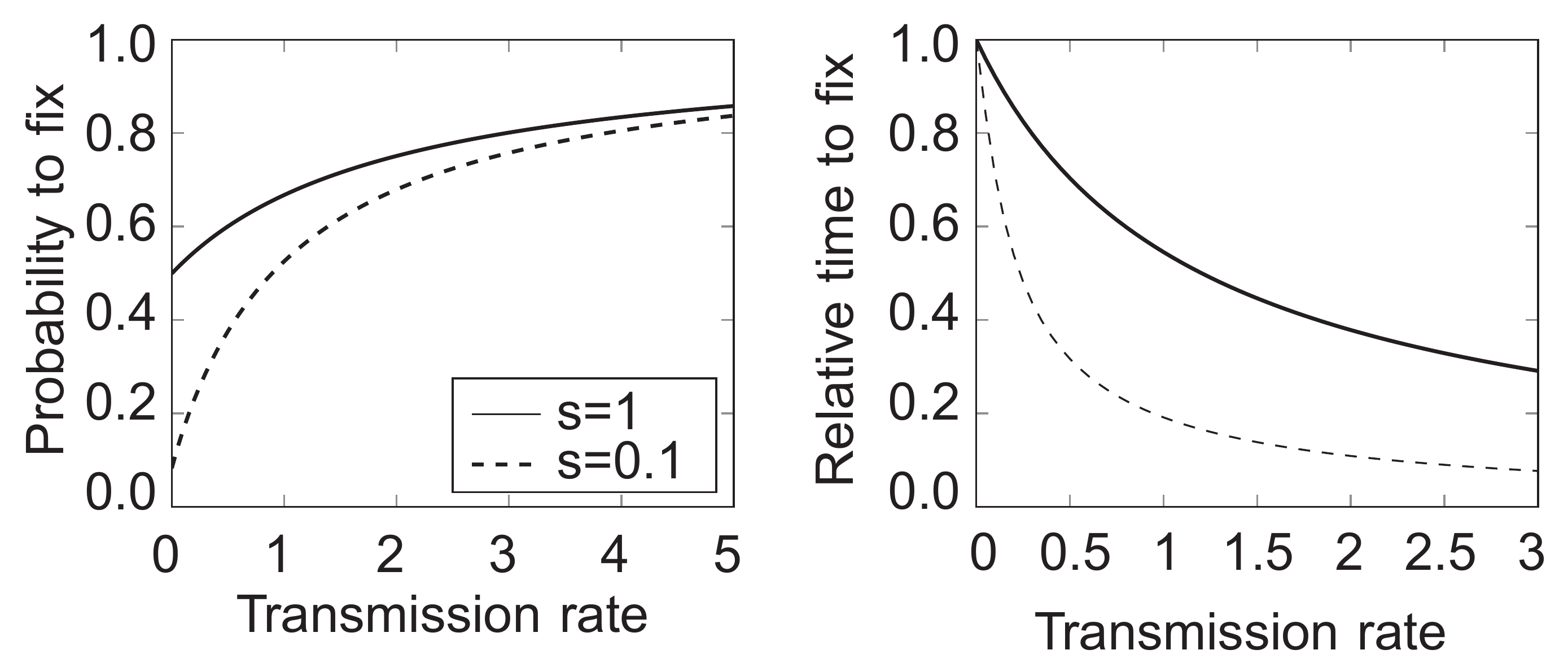}
	\caption{\label{fig:fixed} \textbf{Probability of fixation in the evolutionary model.} The probability that a novel bene fixes (completely saturates the population) as a function of the infection rate, $\beta$, when $s = 1$ and $s = 0.1$ (left panel).  The time to reach a stationary state (where the bene either fixes or goes extinct) as a function of $\beta$, when $s = 1$ and $s = 0.1$ (right panel). Time is plotted relative to the time it takes a match model where $\beta = 0$. $N = 100$ in all cases.}
\end{figure}

Figure \ref{fig:fixed} shows the fixation probability and time to stationarity as a function of $\beta$ for two example values of $s$.
As $\beta$ increases, the probability that the bene fixes increases (left panel). Higher transmission rates also lead to faster time to a stationary solution when compared to those achieved by selection alone (right panel). The decrease in time to reach a stationary solution is more dramatic when selection ($s=0.1$) is lower. Thus for more modest fitness values of the bene the role of horizontal gene transfer is greater.

\section{Epidemics with connectivity benefit}

\subsection{Dynamical equations}
In the dynamical system describing the spread of a bene with connectivity benefits, we use $S$ and $I$ to denote the fraction of nodes
susceptible and infected, respectively, at a given time.
We use $[SI]$ to denote the number of edges between $S$ and $I$ nodes normalized by the total population size,
and so on for $[SS]$ and $[II]$. 
We also define such variables for triplets, such that, for example, $[ISI]$ is the number of node triplets,
such that one of which is susceptible and has edges connecting it to the other two, which are infected.

In the limit of a large population, the change in $S$, $I$, $[SS]$, $[SI]$, and $[II]$ over time is given
by ordinary differential equations
\begin{align}\label{eq:ode-trip}
\dot{I}&=-\dot{S}= \beta [SI] -rI \notag\\
\dot{[SS]}&= -\beta \,[SSI]+r [SI] \frac{k_I-\Delta}{k_I} \notag\\
\dot{[SI]}&= \beta[SSI]  -2\beta[ISI] -\beta[SI] -r[SI] +\\
&\quad\beta[SI]\Delta \frac{S}{S+I\alpha} +2\,r\,[II] \frac{k_I-\Delta}{k_I}  \notag\\
\dot{[II]}&= 2\beta[ISI] + \beta[SI]+\beta\,[SI]\, \Delta \frac{I\alpha}{S+I\alpha} -2r [II] \text, \notag
\end{align}
where $k_I = (2[II]+[SI])/I$ is the average degree of an infected node.

\subsection{Moment closure}
To solve the dynamical equations \eqref{eq:ode-trip}, we need to determine the triplet densities $[ISI]$ and $[SSI]$.
We could write down differential equations for their evolution, but those would involve still new terms specifying
the density of four-node motifs. Therefore, we use a moment-closure approximation to express these triplets in terms of the previously defined pairs.
We do this by assuming that if a node of type $X$ has $k$ incident edges, each of those edges is independently taken to be an $[XY]$ node with
probability proportional to $[XY]$ if $Y\neq X$ and $2[XX]$ if $Y=X$. Therefore, the concentrations of triplets that feature in the
differential equations are given by
\begin{equation}\label{eq:isi}
    \begin{aligned}
	{}[ISI] &= \gamma_S\frac{[SI]^2}{2S}\\
	[SSI] &= \gamma_S\frac{2[SS][SI]}{S}\\
    \gamma_S &= \frac{\langle k^2\rangle_S - \langle k\rangle_S}{\langle k\rangle_S^2}
    \end{aligned}
\end{equation}
The factor $\gamma_S$ compensates for the excess degree of an $S$ node, taking into 
account the fact that the expected number of additional edges a node has conditioned on having at least one edge
is not necessarily the same as the unconditional excepted number of edges. For a Poisson degree distribution,
they are the same, and that factor is $1$. Because, in our graph, edges are continually being created and destroyed,
$\gamma_S$ would also be changing over time, not only because the average degree $\langle k\rangle_S$ would be changing,
but also because the continual redistribution of edges would drive the network toward a Poissonian degree distribution.
To avoid tracking the changes in the degree distribution,
we assume a Poissonian distribution for the susceptible nodes at all times, $\gamma_S=1$.

\subsection{Outbreak dynamics}
Plugging the moment closure \eqref{eq:isi} into the
dynamical equations \eqref{eq:ode-trip}, we obtain a system of four coupled differential equation in four
variables, $I(t)$, $[SS](t)$, $[SI](t)$, and $[SS](t)$, and with parameters denoted by $k_0$, $\Delta$,
$\beta$, $r$, and $\alpha$. We note that the equations satisfy the conservation law
\begin{equation}
    \dot{[SI]}+\dot{[SS]}+\dot{[II]}-\Delta\dot{I} = 0\text,
\end{equation}
which reflects the fact that the total number of edges in the network is directly related to the number
of infected nodes, since each infection event introduces $\Delta$ edges, and each recovery removes the
same number. Therefore, we can eliminate $[II]$ from the system of equations, replacing it with
$[II]=\tfrac12 k_0+\Delta I-[SS]-[SI]$, and be left with three coupled ODEs for three variables.

The state where there are no infected nodes is a fixed point of the system, given by $I=0$, $[SI]=0$,
and $[SS]=\tfrac12 k_0$. To determine whether this fixed point is stable, that is, if a small infection
spreads as an epidemic or dies out, we would normally look at the Jacobian of the system of ODEs at
the fixed point. However, the Jacobian is singular at this fixed point (note that $k_I$ is not well defined
when $I=[SI]=[II]=0$). In order to properly
analyze the stability, we first have to perform a change of variables that resolves the singularity.
One change of variable that accomplishes this task is
\begin{equation}
    \begin{aligned}
	z_1 = \frac{I}{k_0 + 2\Delta I - [SI] - 2[SS]}\\
	z_2 = k_0 + 2\Delta I - [SI] - 2[SS]\\
	z_3 = \frac{\tfrac12 k_0 + \Delta I - [SI] - [SS]}{k_0 + 2\Delta I - [SI] - 2[SS]}\text.
    \end{aligned}
\end{equation}
The values of the old variables at the fixed point, $I=[SI]=[SS]-\tfrac12 k_0=0$, do not
determine the values of $z_1$ and $z_3$. So, we solve the equations $\dot{z}_1=\dot{z}_3=0$
to determine the values of $z_1$ and $z_3$ at the fixed point.

The Jacobian of the time derivatives, $\dot{z}_1$, $\dot{z}_2$, and $\dot{z}_3$, is generically nonsingular at this
fixed point, and its eigenvalues determine the stability of the fixed point. If all eigenvalues
are negative, the fixed point is stable. If any eigenvalue is positive, the fixed point
is unstable. For any value of the parameters
$k_0$, $\Delta$, $r$, and $\alpha$ there is a critical value of the transmissibility $\beta_c$,
such that, if $\beta<\beta_c$, the $I=0$ fixed point is stable, and if $\beta>\beta_c$, the
$I=0$ fixed point is unstable. We find the critical value to be
$\beta_c=r/\left(\tau+\sqrt{k_0+\tau^2}\right)$, where $\tau=\tfrac12(k_0+\delta-1)$.
At this value of $\beta$, the infection free fixed point is given by $z_1=\beta_c/(1+\beta_c)$,
$z_2=0$, and $z_3=\beta_c/2(1+\beta_c)$. The Jacobian can be directly calculated and shown
to have two negative eigenvalues and one zero eigenvalue, as expected.

In the case of a contagion without connectivity benefit, i.e. $\Delta = 0$, we recover the classic SIS dynamics and
$\beta_c = r/k_0$ [31,32]. However, our result for $\beta_c$ is not merely the critical transmission rate for a network with Poisson degree distribution of average $k_0 + \Delta$. On the one hand, the degree distribution of infectious individuals is truncated at values below $\Delta$; it is therefore not exactly Poisson for small $k_0$ and this tends to increase $\beta_c$. On the other hand, and more interestingly, the expected degree of susceptible nodes are also growing with the number of infectious nodes. This last detail is crucial: there is a feedback between the expected epidemic size and the connectivity of the network which lowers the epidemic threshold $\beta_c$.

\subsection{Steady-state Convergence}
When $\beta>\beta_c$, a new stable fixed point emerges with $I>0$. This represents the stable steady-state value of the infected population, where
infections and recoveries occur at the same rate. This value can be obtained by solving the algebraic set of equations given by setting
$\dot I = \dot{[SI]} = \dot{[SS]} = 0$.

\begin{figure}
	\centering
	\includegraphics[width=0.4\linewidth]{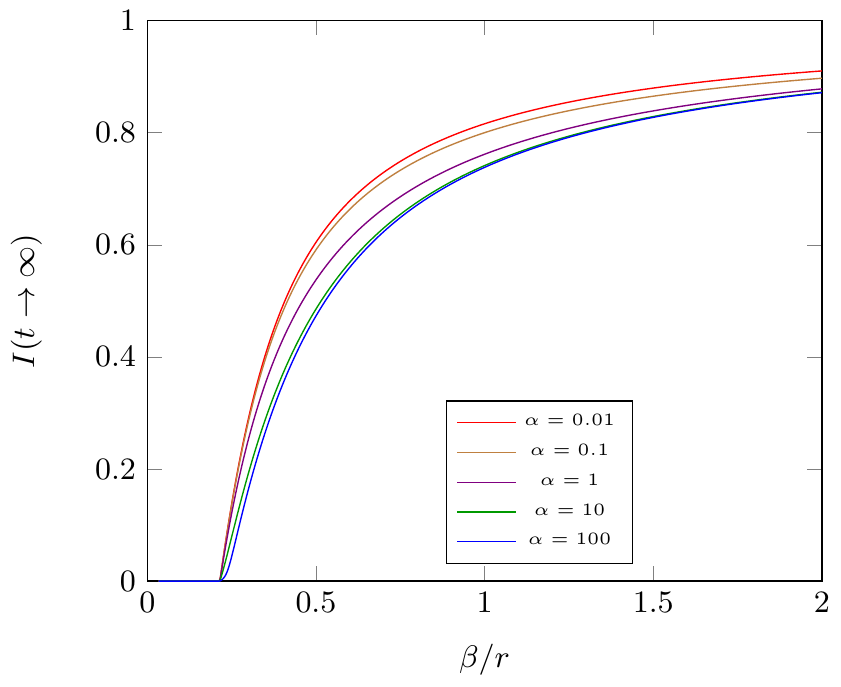} 
	\caption{\textbf{Steady state size of the infected population in the case of instantaneous link addition.} The figure shows the value of the equilibrium value of $I$ as a function of $\beta/r$ when $k_0=3$ and $\Delta=2$. When $\beta<\beta_c$ (here $\beta_c = r/5$), the only possible equilibrium is $I=0$. When $\beta>\beta_c$, the equilibrium state $I=0$ becomes unstable, and a steady equilibrium with $I>0$ emerges.}
	\label{fig:steadystate}
\end{figure}

Figure \ref{fig:steadystate} shows the steady-state fraction of infected individuals as a function of $\beta$. As $\beta$ increases, so does the long-term percentage of infected individuals. The figure also depicts the effect of $\alpha$ on the long term percentage. As $\alpha$ decreases, the targeting of infected individuals improves, and thus the fraction of infected individuals increases.

\subsection{Fixation dynamics under perfect targeting}
We now consider the dynamics at the conclusion of the epidemic. If we consider the case with no recovery ($r = 0$), then the susceptible population always tends to decrease. The rate of this decrease varies with the value of the assortative bias $\alpha$ and the number of new links generated per infection $\Delta$. 

The governing equations in Eq. 4 become simpler in the case where $S \ll 1$ with no recovery. The term $[SS]$ becomes negligibly small as it is second order in the number of $S$ nodes (it requires two S nodes to be connected to one another). The final term in the $\dot{[SI]}$ equation is
\begin{equation}
\label{TheFinalTerm}
\beta [SI] \frac{S}{S+I\alpha}
\end{equation}
\eqref{TheFinalTerm} has two regimes with qualitatively different behavior: $\alpha > 0 $ and $\alpha = 0$.

\eqref{TheFinalTerm} becomes $1$ for $S \ll I\alpha$. The full system thus becomes

\begin{equation}
\label{SIdot}
\begin{aligned}
\dot{S} &= -\beta[SI]\\
\dot{[SI]} &= \beta[SI]\left(-\frac{[SI]}{S} -1\right) + \beta [SI] \Delta\text.
\end{aligned}
\end{equation}
The solution of these coupled ODEs can be seen with a variable substitution called $x$.
\begin{equation}
x = \frac{[SI]}{S}
\end{equation}
With this definition, we get the following relation:
\begin{equation}
\dot{x} = \frac{\dot{[SI]}}{S} - \frac{[SI]}{S}\frac{\dot{S}}{S}
\end{equation}

By substituting the evolution equations for $\dot{S}$ and $\dot{[SI]}$, we get an uncoupled equation. 
\begin{equation}
\dot{x} = \beta x(-x-1+\Delta) - x(-\beta x) = (\Delta - 1) \beta x
\end{equation}
Using Eqn~\ref{SIdot}, we get the system of equations:
\begin{equation}
\begin{aligned}
\dot{x} &= (\Delta - 1) \beta x \\
\dot{S} &= -\beta x S
\end{aligned}
\end{equation}
In this coupled set of equations, $\log x$ changes at a rate $(\Delta -1)\beta$, and $\log s$ changes at a rate of $\beta x$.
\begin{equation}
\label{SofT}
\begin{aligned}
x(t) &\sim \exp[(\Delta -1)\beta t] \\
S(t) &= \exp\left(-\beta \int_0^t x(t')dt'\right)
\end{aligned}
\end{equation}

From \eqref{SofT}, we see that if $\Delta < 1$, the proportion of $S$ nodes decays at an exponential rate that is decaying exponentially. That is, as $S$ decreases, the rate at which $S$ decreases gets slower and slower. Interestingly, because of the exponentially decreasing rate, even as $t \to \infty$ there are always individuals who are not infected.

On the other hand, if $\Delta > 1$ the rate at which $S$ decreases grows in time. In this regime, $S$ never fully reaches $0$, but it tends to 0 more and more quickly as the epidemic spreads. At the critical point, $\Delta = 1$, the rate at which $S$ shrinks is constant. 

\subsection{Fixation dynamics under imperfect targeting}

If $\alpha > 0$, $\frac{S}{S+I\alpha}$ approaches $\frac{S}{\alpha}$ as $S$ approaches 0. The coupled ODE system reduces to
\begin{equation}
\begin{aligned}
\dot{S} &= -\beta[SI]\\
\dot{[SI]} &= \beta[SI]\left(-\frac{[SI]}{S} -1\right) + \beta [SI] \Delta \frac{S}{\alpha} \text.
\end{aligned}
\end{equation}
Here, there are two regimes:
\begin{equation}
\begin{aligned}
I \gg S \gg \alpha \\
\alpha \gg S
\end{aligned}
\end{equation}

In the first regime where $I \gg S \gg \alpha$, $\frac{S}{S+I \alpha}$ is approximately 1 which means that the behavior is the same as if there is perfect targeting ($\alpha = 0$). However, as the infection proceeds and $S$ gets sufficiently small, the regime switches. As a result $\frac{S}{S+I \alpha}$ is approximately 0, and so the behavior is as if $\Delta = 0$. As $S$ becomes very small, new links to susceptible individuals are added with increasing low frequency. Thus, in the final stages of the epidemic, the additional links added by newly infected individuals only have an impact if they perfectly attach to susceptible individuals.

\subsection{Continuous link creation}

We now consider a case where the extra connectivity accrues throughout
the time an individual is infected.
To keep the analysis simple, we ignore the possibility that an infected node recovers. The system of differential equations describing the system is
\begin{equation} \label{eq:ode-cont}
    \begin{aligned}
\dot{I}&=-\dot{S}= \beta [SI] \\
\dot{[SS]}&= -\beta \,[SI]\, 2\frac{[SS]}{S} \\
\dot{[SI]}&= \beta[SI] \left( 2\frac{[SS]}{S} -\frac{[SI]}{S}-1 \right) + I \Delta \frac{S}{S+I\alpha} \\
\dot{[II]}&= \beta[SI] \left( \frac{[SI]}{S} +1 \right) +I \Delta \frac{I\alpha}{S+I\alpha}
    \end{aligned}
\end{equation}

\subsection{Outbreak dynamics}
In the continuous link-addition model \eqref{eq:ode-cont},
the spread of the epidemic accelerates due to the continued increase of the degree of infected nodes. To determine
the outbreak spreading rates, we consider the equations for $\dot I$ and $\dot{[SI]}$ shown in \eqref{eq:ode-cont}.
When $I\ll 1$, the nonnegligible terms are
\begin{equation} \label{eq:ode-cont-gen}
    \begin{aligned}
\dot{I} &= \beta [SI] \\
\dot{[SI]}&= \beta[SI] \left( k_0 - 1 \right) + I \Delta \text.
    \end{aligned}
\end{equation}
This coupled system of ordinary differential equations can be rewritten using the compound variable $\mathbf{y} = (I, [SI])^T$, giving the simple equation $\dot{\mathbf{y}} = \mathbf{A}\mathbf{y}$, where
\begin{equation}
\mathbf{A} = \left(\begin{array}{cc} 0 & \beta \\ \Delta & \beta (k_0 -1) \end{array}\right)
\end{equation}
The eigenvalues of $\mathbf{A}$ are $\lambda_\pm=\tfrac12 \left[\beta(k_0-1) \pm (4\Delta\beta+\beta^2(k_0-1)^2)^{1/2}\right]$. At long times, both $I$ and $[SI]$ grow
exponentially as $\exp(\lambda_+ t)$, and the time scale for this behavior to take
hold is $1/(\lambda_+-\lambda_-)$. 

Even though infected individuals keep acquiring new
links and their degree grows without bound, the rate of growth of the epidemics saturates
at $\lambda_+$ because the newly-infected individuals start with the background number of neighbors, $k_0$. Therefore, the degree of the typical infected individual will grow to a steady state value in the exponential growth phase of the epidemic.
\subsection{Fixation dynamics}
To analyze fixation dynamics we focus on the following equations in which we assume that $[SS]/S$ is negligible: 
\begin{equation}\label{eq:fix-cont}
\begin{aligned}
\dot{S} &= -\beta[SI]\\
\dot{[SI]} &= -\beta[SI]\left(\frac{[SI]}{S} +1\right) + \Delta \frac{S}{S+I\alpha}\text.
\end{aligned}
\end{equation}

If targeting is imperfect ($\alpha > 0$), when $S$ becomes small enough that $S<\alpha$, the last
term in \eqref{eq:fix-cont} will be approximately equal to $\Delta S/\alpha$. Using this substitution, we get the following equation for the time evolution of the variable $x = [SI]/S$:
\begin{equation}
\dot{x} = -\beta x + \frac{\Delta}{\alpha}\text.
\end{equation}
This will eventually saturate to the steady state value $x^{*} = \Delta/(\alpha\beta)$,
and the susceptible rate, governed by the equation $\dot{S} = -\beta x S$, will decay exponentially as $\exp(-\beta x^{*} t)=\exp(-\Delta \cdot t/\alpha)$. In contrast to the instantaneous link-addition model, the rate does not decay exponentially. Thus, the fraction of $S$ decreases faster and as $t \to \infty$, $S \to 0$.

However, if $\alpha=0$ (perfect targeting) then the last term of \eqref{eq:fix-cont} is simply $\Delta$. Unlike all other cases, the rate of susceptible individuals will become zero at a finite time. To see why this behavior is the solution to
the differential equations when $S$ approaches zero, we use the following \textit{ansatz}:
\begin{equation}\label{eq:ansatz}
\begin{aligned}
S(t) &= S_0 (t^*-t)^a \\
[SI](t) &= [SI]_0 (t^*-t)^b \text.
\end{aligned}
\end{equation}
The first equation of \eqref{eq:fix-cont} gives 
\begin{equation}\label{eq:plug-ansatz1}
-a S_0 (t^*-t)^{a-1} = -\beta[SI]_0 (t^*-t)^b\text{,}
\end{equation}
implying that $a = b+1$. This also implies that near $t^*$, $[SI]/S\gg 1$, and therefore,
the second equation of \eqref{eq:fix-cont} gives
\begin{equation}\label{eq:plug-ansatz2}
-b [SI]_0 (t^*-t)^{b-1} = -\beta \frac{[SI]_0^2}{S_0}(t^*-t)^{b-1}+\Delta\text.
\end{equation}
For both right-hand-side terms to be comparable, we need $b=1$. Finally, we recover
the prefactors $S_0=\tfrac12 \beta\Delta$ and $[SI]_0=\Delta$ from \eqref{eq:plug-ansatz1}
and \eqref{eq:plug-ansatz2}.

Interestingly, if $\alpha$ is small but nonzero, then as in the instantaneous link-adding scenario, the dynamics of the infection will cross over from a regime that behaves as if $\alpha=0$, that is where $S$ approaches zero quickly and appears headed to vanish at some finite time, to, once $S$ becomes comparable to $\alpha$, a regime of regular exponential decay with rate constant $\Delta/\alpha$.

Another interesting consequence is the difference in the continuous case between perfect targeting and even slightly flawed targeting ($\alpha > 0$). If the targeting is perfect, then all of the individuals will be infected in finite time for \emph{any} positive $\Delta$, the number of links created per unit time. In constrast, no matter how large the value of $\Delta$, if $\alpha$ is non-zero, it still takes an infinite amount of time to infect all individuals. Error-prone infected individuals cannot convert the whole population in finite time as eventually the false-positive identifications dominate the true-positive ones as the fraction of susceptible individuals becomes increasingly small. Alternately, converting the entire population in finite time requires each infected individual to create a non-zero number of new links with susceptible individuals on average per time step. Of course, the time for the number of susceptible individuals to reach a small fraction of the population ($S < \epsilon$) will depend greatly on the link generation rate for either $\alpha$ regime.

\section{Epidemics with utility benefit}\label{app:social}
\subsection{General Dynamics}
In the main text, we showed how explicit preferences can lead to strategic assortative or disassortative rewiring. From this, we can derive the rates of change of $I$, $S$, $[SI]$, $[II]$ and $[SS]$. $[SI]$, $[II]$ and $[SS]$ are normalized to sum to the average number of edges per individual $E/N$.

Consider a population of size $N$ with $E$ edges. We assume a well-mixed population.

The infection dynamics without rewiring are the same as in the connectivity benefits model, without recovery $r=0$ or connectivity benefit $\Delta=0$ (see Eq. 4 in the main text):
\begin{align*}
\dot{I}&= \beta [SI] \\ 
\dot{[SI]}&= \beta[SI] \left( 2\frac{[SS]}{S} -\frac{[SI]}{S}-1 \right) \\
\dot{[II]}&= \beta[SI] \left(\frac{[SI]}{S} +1\right)
\end{align*}
and $S=1-I$, $[SS] = ~\frac{E}{N} -[SI] - [II]$.

Infected individuals that rewire disassortatively do so at a rate:
\[
r_{i-i\to s} = \frac{I}{E}(1-(1-\alpha_{II})^{k_I})
\]
This is the proportion of infected individuals that have at least one II link they wish to replace by an SI link. Here $\alpha_{II} = \frac{2 [II]}{[SI]+2[II]}$ is the probability that a stub coming out of an infected node is an II stub, and $k_I = E\frac{ [SI] + 2 [II]}{NI}$ is the average degree of infected nodes.  

Infected individuals that rewire assortatively do so at a rate:
\begin{align*}
r_{i-s\to i} = \frac{I}{E}(1-(1-\alpha_{SI}^I)^{k_I})
\end{align*}
$\alpha_{SI}^I = \frac{[SI]}{[SI]+2[II]}$ is the probability that a stub coming out of an infected node is an SI stub.  

As we saw, for susceptibles, assortative and disassortative rewiring rates can depend on the number of infected neighbors.  We use Eq. 6 and compute the probability that a node $S$ meets the condition for disassortative rewiring assuming $S_n = k_S - I_n$ where $k_S \equiv E\frac{ [SI] + 2 [SS]}{NS}$ is the average degree of susceptible nodes.  Susceptibles prefer assortative rewiring as long as $I_n < \hat{I}_n$, where 
$$\hat{I}_n = \frac{-d_S/\beta - c_I + (d_I-d_S) - (b_I-b_S) k_S - (c_I-c_S) N I}{2(d_I-d_S)}$$ 
Therefore, the rewiring rates for susceptibles:
\begin{align*}
r_{s-i \to s} & = \frac{S}{E} P(1\leq {I}_n \leq \hat{I}_n ) = \frac{S}{E} \sum_{j=1}^{\hat{I}_n} {\binom{k_S}{j}} (\alpha_{SI}^S)^{j}(1-\alpha_{SI}^S)^{k_S-j}\\
r_{s-s \to i} & = \frac{S}{E}  P(1\leq {S}_n \leq k_S-\hat{I}_n) = \frac{S}{E}  \sum_{j=1}^{k_S-\hat{I}_n} {\binom{k_S}{j}}(\alpha_{SS})^{j}(1-\alpha_{SS})^{k_S-j}
\end{align*}
where $\alpha_{SS}= \frac{2[SS]}{[SI]+2[SS]}$ and $\alpha_{SI}^S = \frac{[SI]}{[SI]+2[SS]}$. 

We use these rates and the analysis of predicted utility in the main text to derive dynamics corresponding to the three cases considers: \textit{evangelizers}, \textit{cool kids}, and \textit{snobs}.

\subsection{Evangelizers case}

The dynamics in the evangelizers case is given by the following ODEs:
\begin{align*}
\dot{I} &= \beta [SI]\\
\dot{[SI]} &= \beta[SI] \left( 2\frac{[SS]}{S} -\frac{[SI]}{S}-1 \right) + \frac{I}{E}(1-(1-\alpha_{II}^I)^{k_I}) + \frac{S}{E}(1-(1-\alpha_{SS})^{k_S})\\
\dot{[II]} &=\beta[SI] \left(\frac{[SI]}{S} +1\right) - \frac{I}{E}(1-(1-\alpha_{SI}^I)^{k_I})
\end{align*}
To examine the breakout dynamics, consider the initial situation where $I\ll 1$, $[SI]^2 \approx 0$, $[SS]/S \approx 1$, $\alpha_{SS} \approx 1$, $\alpha_{II}^I \approx 0$. We can simplify this system to:
\begin{align*}
\dot{I} &= \beta [SI]\\
\dot{[SI]} &= \beta [SI] + 1/E
\end{align*}
Initial growth is thus exponential.

To examine the fixation dynamics, consider the limiting situation where $S \ll 1$, $[SS] \approx 0$, $\alpha_{SS} \approx 0$, $\alpha_{II}^I \approx 0$. We can now simplify this system to:

\begin{align*}
\dot{I} &= \beta [SI]\\
\dot{[SI]} &= -\beta [SI]\left(\frac{[SI]}{S} + 1 \right) + I/E
\end{align*}
These equations take the same form as \eqref{eq:fix-cont} when $\alpha = 0$ (continuous link addition with perfect targeting), which was shown to correspond to super-exponential fixation.

\subsection{Cool kids case}
The cool kids case gives the following ODEs:
\begin{align*}
\dot{I} &= \beta [SI]\\
\dot{[SI]} &= \beta[SI] \left( 2\frac{[SS]}{S} -\frac{[SI]}{S}-1 \right) -\frac{I}{E} (1-(1-\alpha_{SI}^I)^{k_I}) + \frac{S}{E} (1-(1-\alpha_{SS})^{k_S})\\
\dot{[II]} &=\beta[SI] \left(\frac{[SI]}{S} +1\right) + \frac{I}{E}(1-(1-\alpha_{SI}^I)^{k_I})
\end{align*}

\subsection{Snobs case}
In the snobs case $\hat{I}_n = \frac{S_n + 2}{3} = \frac{k_S+2}{4}$. The dynamics are described by the following equations:
\begin{align*}
\dot{I} &= \beta [SI]\\
\dot{[SI]} &= \beta[SI] \left( 2\frac{[SS]}{S} -\frac{[SI]}{S}-1 \right) -\frac{I}{E} (1-(1-\alpha_{SI}^I)^{k_I}) \\
&\;\;\;\;\; - \frac{S}{E}  \sum_{j=1}^{(k_S+2)/4} {\binom{k_S}{j}} (\alpha_{SI}^S)^j (1-\alpha_{SI}^S)^{k_S-j} + \frac{S}{E}  \sum_{j=1}^{k_S - (k_S+2)/4} {\binom{k_S}{j}}(\alpha_{SS})^j(1-\alpha_{SS})^{k_S-j}
\\
\dot{[II]} &=\beta[SI] \left(\frac{[SI]}{S} +1\right) +\frac{I}{E} (1-(1-\alpha_{SI}^I)^{k_I})
\end{align*}

\subsection{Agent-based version: robustness of mean-field results to stochasticity and heterogeneity}
The ODE model can be seen as a mean-field approximation of a discrete stochastic epidemic-spreading process, one with local differences in connectivity, degree heterogeneity, and correlated network properties.  In order to check the robustness of the mean-field approximation, here we present simulation results of a stochastic discrete agent-based version of the same basic model.

\begin{figure}
\centering
\includegraphics[width=.9\linewidth]{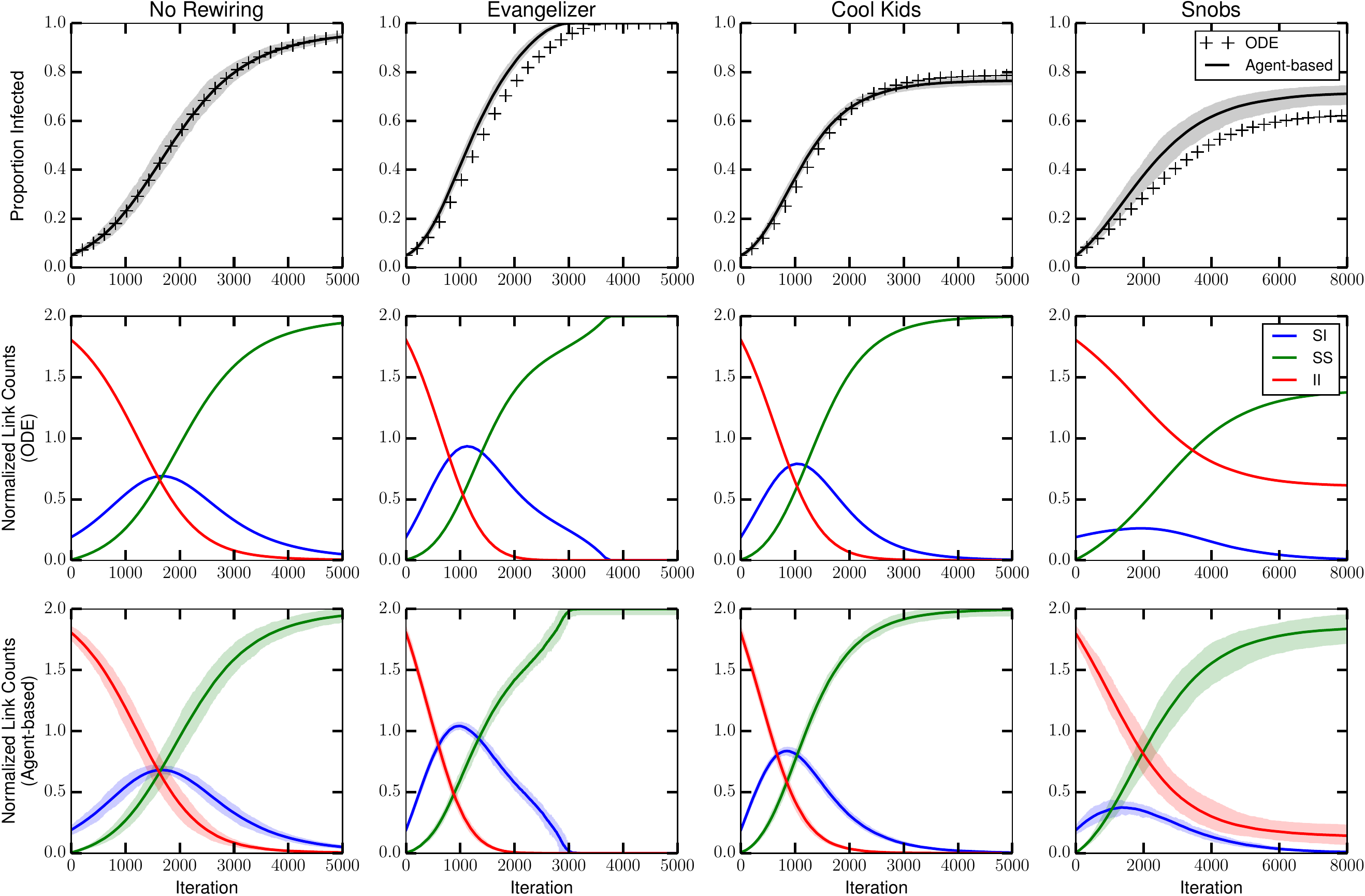}
\caption{Comparison between the agent-based model and the mean-field approximation (ODE model) presented in the main text. For both models, parameters are $N=1000$, $E=4000$, initial infection rate $0.05$, $\beta=5\times 10^{-4}$. The solid line indicates the mean, and the shaded area represents the 10\% to 90\% percentiles across 100 runs.}\label{fig:ABM_SM}
\end{figure}

In the agent-based model, agents have the same utility function as described in the main text. $N$ agents are distributed on a network with $E$ edges, which starts off as Erd\H{o}s-R\'{e}nyi random graph. At each time step, a randomly selected individual first attempts to adaptively rewire, following equations Eq. 5 and Eq. 6 according to its state and neighborhood. Afterward, each infected-susceptible edge leads to the spread of an infection with probability  $\beta$. Figure \ref{fig:ABM_SM} compared the results of the ODE model with the results of the agent-based simulations, for each of the three cases defined in the main text as well as a no-rewiring condition.  Parameters are set same as in the main text (population of $N=1000$ agents, $E=4000$ edges, initial infection rate $0.05$, $\beta=5\times 10^{-4}$). We see that the dynamics are qualitatively the same in both the discrete agent-based version and the continuous mean-field version.

\end{document}